\documentclass[aps,prd,preprintnumbers,superscriptaddress,nofootinbib]{revtex4}%
\usepackage[dvipdfmx]{graphicx} 
\usepackage{feynmf}
\usepackage{color}
\usepackage{bm,latexsym,amsmath,amssymb,amsfonts,mathrsfs}
\usepackage[colorlinks=true,linkcolor=magenta,citecolor=magenta]{hyperref}

\usepackage{physics}	
\usepackage{xcolor}
\usepackage{tensor}
\usepackage{ulem}

\usepackage{physics}
\usepackage{aas_macros}

\newcommand*{\beqa}{\begin{eqnarray}}
\newcommand*{\eeqa}{\end{eqnarray}}

\newcommand*{\p}{\partial}
\newcommand{\realR}{\mathbb{R}}
\newcommand{\ra}{\rangle}
\newcommand{\la}{\langle}
\renewcommand{\H}{{\cal H}}

\renewcommand{\P}{{\bf P}}
\newcommand{\x}{{\bf x}}

\begin{document}

\title{Quantum Clocks, Gravitational Time Dilation, and Quantum Interference}

\author{
Takeshi Chiba
}
\affiliation{Department of Physics, College of Humanities and Sciences, Nihon University, \\
                Tokyo 156-8550, Japan}
\author{
Shunichiro Kinoshita
}
\affiliation{Department of Physics, College of Humanities and Sciences, Nihon University, \\
                Tokyo 156-8550, Japan}
\affiliation{
Department of Physics, Chuo University, 
Kasuga, Bunkyo-ku, Tokyo 112-8551, Japan
}
\date{\today}

\begin{abstract}
A proper time observable for a quantum clock is introduced and it is found that 
the proper time read by one clock conditioned on another clock  
 reading a different proper time obeys classical time dilation in accordance 
 with special relativistic kinematical time dilation. 
 Here, we extend this proposal to a weak gravitational field 
in order to investigate whether the weak equivalence principle holds for 
quantum matter. 
We find that for general quantum states 
the quantum time dilation in a weak gravitational field obeys 
a similar gravitational time dilation found in classical relativity. 
However, unlike the special relativistic case, the time dilation involves 
the external time (a background coordinate time at the observer on the Earth) as well as 
the proper times of two clocks. We also investigate a quantum time dilation effect 
induced by a clock in a superposition of wave packets localized in momentum space or in position space and 
 propose the setup to observe the gravitational effect in 
 the quantum interference effect in the time dilation. 
\end{abstract}


\maketitle

\section{Introduction}

The equivalence principle played the fundamental role in constructing 
general relativity (GR) by Einstein.  
The weak equivalence principle (WEP) states that 
the motion of a (uncharged) test body is independent of its internal structure 
and composition \cite{will2018}. 
With the WEP together with the local Lorentz invariance 
(independence of the results of nongravitational experiments from 
the velocity of the local Lorentz frame) 
and the local position invariance 
(the independence of the experimental results from the spacetime position),  
the metric couples universally to matter. 
However, since this ``universal coupling to matter'' is the root of the problem 
of the cosmological constant, the validity of the WEP should be examined 
further,  especially in the quantum regime. 
Tests of the weak equivalence principle have been performed 
by measuring relative acceleration between 
different atoms using atom interferometers \cite{1999Natur.400..849P,Fray:2004zs,2013PhRvA..88d3615B,Schlippert:2014xla,Albers:2020fag,Asenbaum:2020era,Barrett:2021cdy}. 

There are several attempts to formulate the equivalence principle 
in the quantum regime \cite{Zych:2015fka,Anastopoulos:2017jik,Hardy:2019cef,Roura:2018cfg,Roura:2020qbu,Giacomini:2021aof,Cepollaro:2021ccc}. 
In addition to the position of a particle which is measured by interferometric experiments, an important observable quantity 
regarding the equivalence principle 
is the proper time of a clock 
because the gravitational time dilation is the direct consequence of 
the equivalence principle \cite{will2018}. 
Measurements of  the gravitational time dilation by comparisons of atomic clocks 
located at different heights have been performed  
 \cite{Chou:2010zz,Takamoto:2020tux,Bothwell:2021fqe,Zheng:2022hwj}. 

Recently, a proper time observable for a quantum clock is proposed 
in \cite{Smith:2019imm} and the probability that one clock reads a given proper 
time conditioned on another clock reading a different proper time is derived. 
Moreover, it is shown that when the external degrees of freedom of these clock particles 
are described by Gaussian wave packets localized in momentum space, 
the clocks observe  classical time dilation in accordance 
 with special relativistic kinematical time dilation.

In this paper, we extend this proposal to a weak gravitational field 
and study the time dilation of quantum particles 
under the Newtonian gravity in order to investigate  whether quantum particles 
move in the same manner as the classical particles. 
Moreover, we are able to derive the quantum time dilation between two clocks 
for general quantum states not limited to Gaussian wave packets. 
Thanks to the existence of the conserved quantities associated with 
the Galilean invariance and the translation invariance of the system, 
it can be shown that the gravitational time dilation depends only on 
the expectation values of the positions and the velocities of clocks at the initial time.  
 This may be regarded as the  weak equivalence principle for quantum system \cite{Anastopoulos:2017jik}.

The paper is organized as follows. In Sec. \ref{sec2}, we derive 
the quantum time dilation formula in a weak gravitational field. 
In Sec. \ref{sec3},  we calculate the gravitational time dilation 
for a linear gravitational potential.  We also consider a quantum time dilation effect 
induced by a clock in a superposition of wave packets. 
Sec. \ref{sec4} is devoted to summary.  In Appendix \ref{app:Gaussian}, the time dilation 
for Gaussina wavepackets is given,  and in Appendix \ref{app:superposition}, the initial expectation values for a superposition states are given.

\section{Quantum Clock Particles in Spacetime}
\label{sec2}

\subsection{Classical Particles}

We consider a system of $N$ free particles.  Each particle whose mass is 
 $m_n~(n=1,\dots,N)$ has a set of internal 
degrees of freedom, labeled by the configuration variables $q_n$ and 
their conjugate momenta $P_{q_n}$\cite{Smith:2019imm}. These internal 
degrees of freedom is supposed to represent the quantum clock.

The action of such a system in a curved spacetime with the metric 
$g_{\mu\nu}$ is given by
\beqa
S=\sum_n\int d\tau_n\left(-m_nc^2+P_{q_n}\frac{dq_n}{d\tau_n}-
H_n^{\rm clock}\right) ,
\eeqa
 where $\tau_n$ is the proper time of the $n$th particle and 
 $H_n^{\rm clock}=H_n^{\rm clock}(q_n,P_{q_n})$ is the Hamiltonian 
 for its internal degrees of freedom.  
 
 Let  $x_n^{\mu}$ denote the spacetime position of the $n$th particle. 
 The trajectory of the $n$th particle $x_n^{\mu}(t)$  
 is parametrized by an arbitrary external time parameter $t$.  Noting that 
 $cd\tau_n=\sqrt{-g_{\mu\nu}\dot x_n^{\mu}\dot x_n^{\nu}}dt\equiv\sqrt{-\dot x^2_n}dt$, 
 where a dot denotes differentiation with respect to $t$,  
 the action is rewritten as 
 \beqa
 S=\int dt\sum_n\frac{1}{c}\sqrt{-\dot x_n^2}
 \left(-m_nc^2+P_{q_n}c\frac{\dot q_n}{\sqrt{-\dot x_n^2}}-H_n^{\rm clock}\right) =: \int dt~ L .
 \eeqa
 The momentum conjugate to $x_n^{\mu}$ is given by
 \beqa
 P_{n\mu}=\frac{\p L}{\p \dot x_n^{\mu}}=
 \frac{g_{\mu\nu}\dot x_n^{\nu}}{c\sqrt{-\dot x_n^2}}\left(m_n c^2+H_n^{\rm clock}\right) .
 \eeqa
 Then the Hamiltonian associated with the Lagrangian $L$ is constrained to vanish:
\beqa
H=\sum_n\left(P_{n\mu}\dot x_n^{\mu}+P_{q_n}\dot q_n\right)-L\approx 0 .
\eeqa
In terms of the momentum, the constraints can be expressed in the form
\beqa
C_{H_n}:=g^{\mu\nu}P_{n\mu}P_{n\nu}c^2+\left(m_nc^2+H_n^{\rm clock}\right)^2\approx 0 .
\eeqa
Using the $(3+1)$ decomposition of the metric in terms of the lapse function $\alpha$, 
the shift vector $\beta^i$ and the three-metric $\gamma_{ij}$ such that \cite{mtw} 
\beqa
ds^2=-\alpha^2 c^2dt^2+\gamma_{ij}(dx^i+\beta^icdt)(dx^j+\beta^jcdt),
\eeqa
the constraint is factorized in the form
\beqa
C_{H_n}=-\alpha^{-2}\left(P_{n0}-\beta^iP_{ni}\right)^2c^2+\gamma^{ij}P_{ni}P_{nj}c^2+\left(m_nc^2+H_n^{\rm clock}\right)^2= -\alpha^{-2}C_n^+C_n^-\approx 0 ,
\label{constraint-H}
\eeqa
where $C_n^{\pm}$ is defined by 
\beqa
C_n^{\pm}&:=&\left(P_{n0}-\beta^iP_{ni}\right)c\pm h_n ,
\label{constraint-pm}\\
h_n&:=&\alpha\sqrt{\gamma^{ij}P_{ni}P_{nj}c^2+(m_nc^2+H_n^{\rm clock})^2} .
\label{hamiltonian-n}
\eeqa
Note that we have set $x^0 = ct$.
Hereafter we assume that  the spacetime is static and 
the shift vector is vanishing, $\beta^i=0$.  
The coordinates $x^{\mu}_n$ and their conjugate momenta $P_{n\mu}$ satisfy 
the fundamental Poisson brackets: $\{x_m^{\mu},P_{n\nu}\}=\delta_{mn}\delta^{\mu}_{\nu}$. 
The canonical momentum $P_{n\mu}$ generates translations in the spacetime coordinate 
$x_n^{\mu}$. Therefore, if $C_n^{\pm}\approx 0$ and  the shift vector vanishes, 
then $\pm h_n$ is the generator of translation in the $n$th particle's time coordinate
and is the Hamiltonian for both 
the external and internal degrees of freedom of the $n$th particle.

\subsection{Quantization}

We canonically quantize the system of $N$ particles by promoting the phase space 
variables to operators acting on appropriate Hilbert spaces: 
$x^0_n$ and $P_{n0}$ become canonically conjugate self-adjoint operators acting on the Hilbert space ${\cal H}^0_n\simeq L^2(\mathbb{R})$ associated with 
the $n$th particle's temporal degree of freedom; operators $x_n^i$ and $P_{ni}$ 
acting on the Hilbert space 
${\cal H}_n^{\rm ext}\simeq L^2(\realR^3)$ associated with the particle's 
external degrees of freedom; operators $q_n$ and $P_{q_n}$ acting  
on the Hilbert space 
${\cal H}_n^{\rm clock}$ associated with the particle's 
internal degrees of freedom. Then the Hilbert space describing 
the $n$th particle is 
${\cal H}_n\simeq {\cal H}^0_n\otimes{\cal H}_n^{\rm ext}\otimes{\cal H}_n^{\rm clock}$.

The constraint equations (\ref{constraint-H}) now become operator equations restricting the physical state of the theory,  
\beqa
C_n^+C_n^-|\Psi\rangle\rangle=0,~~~~\forall n ,
\eeqa 
where $|\Psi\rangle\rangle\in {\cal H}_{\rm phys}$ is a physical state of a clock $C$ and a system $S$ and lives in the 
physical Hilbert space ${\cal H}_{\rm phys}$. 

To specify ${\cal H}_{\rm phys}$, following Page and Wooters~\cite{Page:1983uc,1984IJTP...23..701W} (see also \cite{Giovannetti:2015qha,Castro-Ruiz:2019nnl,Maccone:2018cyv}), 
the normalization of the physical state in ${\cal H}_{\rm phys}$ is performed by 
projecting a physical state $|\Psi\rangle\rangle$ onto a subspace in which 
the temporal degree of freedom of each particle (clock $C$) 
is in an eigenstate $|t_n\rangle$ 
of the operator $x_n^0$ associated with the eigenvalue $t\in \realR$ in 
the spectrum of $x_n^0$: $x_n^0|t_n\rangle= c t|t_n\rangle$. 
The state of $S$ by conditioning $|\Psi\ra\ra$ on $C$ reading the time $t$ is then 
given by
\beqa
|\psi_S(t)\ra=\la t|\otimes I_S|\Psi\ra\ra ,
\eeqa
where $|t\ra=\otimes_n|t_n\ra$ and $I_S$ is the identity on 
${\cal H}\simeq \bigotimes_n\H_n^{\rm ext}\otimes\H_n^{\rm clock}$.  
We demand that the state  $|\psi_S(t)\ra$ is normalized as $\la\psi_S(t)|\psi_S(t)\ra=1$ for $\forall t\in \realR$  on 
a spacelike hypersurface defined by all $N$ particles' temporal degree 
of freedom being in the state $|t_n\ra$. The physical state $|\Psi\ra\ra$ 
is thus normalized with respect to the inner product~\cite{Smith:2019imm}:
\beqa
\la\la\Psi|\Psi\ra\ra_{PW}:=\la\la\Psi||t\ra\la t|\otimes I_S|\Psi\ra\ra
=\la\psi_S(t)|\psi_S(t)\ra=1, 
\eeqa
 and the physical state $|\Psi\ra\ra$ can be written as
\beqa
|\Psi\ra\ra=\int dt |t\ra\la t|\otimes I_S|\Psi\ra\ra=\int dt|t\ra|\psi_S(t)\ra .
\label{entangled}
\eeqa

 Hereafter, we consider physical states that satisfy
 \beqa
 C_n^+|\Psi\rangle\rangle=(P_{n0}c+h_n)|\Psi\rangle\rangle=0,~~~~~\forall n,
 \eeqa
 where $h_n$ is the operator corresponding to Eq. (\ref{hamiltonian-n}). 
This implies that the conditioned state $|\psi_S(t)\ra$ has positive energy as 
measured by $h_n$.  
It can be shown  that the conditioned state $|\psi_S(t)\ra$ satisfies the 
Schr\"odinger equation with  $t$ as a time parameter \cite{Smith:2019imm}:
\beqa
i\hbar \frac{d}{dt}|\psi_S(t)\ra=H_S|\psi_S(t)\ra ,
\label{schrodinger}
\eeqa
where $H_S=\sum_nh_n\otimes I_{S-n}$ with $I_{S-n}$ being the identity 
on $\bigotimes_{m\neq n}\H^{\rm ext}_m\otimes\H^{\rm clock}_m$.  
Therefore,   $|\psi_S(t)\ra$ can be regarded as the 
time-dependent state of the $N$-particles with the Hamiltonian $H_S$ 
evolved with the external time $t$.

\subsection{Proper Time Observables}

In \cite{Smith:2019imm}, a clock is defined to be the quadrupole 
$\{\H^{\rm clock},\rho_C,H^{\rm clock}, T_C\}$, 
where $\rho_C$ is a fiducial state and $T_C$ is proper time observable. 
The proper time observable is defined as a POVM (positive operator valued measure)
\beqa
T_C:=\left\{ E_C(\tau)~\forall \tau \in G ~{\rm s.t.}\int_G d\tau E_C(\tau)=I_C\right\} ,
\eeqa
where $E_C(\tau)=|\tau\ra\la\tau|$ is a positive operator on $\H^{\rm clock}$, $G$ is the group generated by $H^{\rm clock}$, and $|\tau\ra$ is a 
clock state associated with a measurement of the clock yielding the time $\tau$. 
The clock state $|\tau\ra$ evolves according to $U_C(\tau)=e^{-iH^{\rm clock}\tau/\hbar}$ as
\beqa
|\tau+\tau'\ra=U_C(\tau')|\tau\ra .
\eeqa 

\subsection{Probabilistic Time Dilation}

Consider two clock particles $A$ and $B$  with internal degrees of freedom, $\{\H^{\rm clock}_A,\rho_A,H^{\rm clock}_A, T_A\}$ and 
 $\{\H^{\rm clock}_B,\rho_B,H^{\rm clock}_B, T_B\}$. 
To probe time dilation effects between two clocks, we consider 
the  probability that clock $A$ reads the proper time $\tau_A$ 
conditioned on clock $B$ reading the proper time $\tau_B$
\cite{Page:1983uc,1984IJTP...23..701W}.
This conditional probability is given in terms of the physical state as
\beqa
{\rm Prob}[T_A=\tau_A|T_B=\tau_B]=\frac{\la\la\Psi|E_A(\tau_A)
E_B(\tau_B)|\Psi\ra\ra}{\la\la\Psi|
E_B(\tau_B)|\Psi\ra\ra} \,.
\label{conditional-prob}
\eeqa

Consider the case where two clock particles $A$ and $B$ are moving in a 
curved spacetime. 
Suppose that initial conditioned state is unentangled, 
$|\psi_S(0)\ra=|\psi_{S_A}\ra |\psi_{S_B}\ra$, and that the external and internal degrees of freedom of both particles are unentangled, 
$|\psi_{S_n}\ra=|\psi_n^{\rm ext}\ra |\psi_n^{\rm clock}\ra$. 
Then, from Eq. (\ref{entangled}), 
the physical state takes the form
\beqa
|\Psi\ra\ra=\int dt |t\ra |\psi_S(t)\ra
=\int dt \bigotimes_{n\in \{A,B\}}e^{-ih_nt/\hbar}|\psi_n^{\rm ext}\ra|\psi_n^{\rm clock}\ra\,.
\eeqa 
Further suppose that $\H^{\rm clock}_n\simeq L^2(\realR)$ so that 
we may consider an ideal clock such that  
 $P_n = H^{\rm clock}_n/c$ and $c T_n$ are 
the momentum and position operators on  $\H^{\rm clock}_n$. 
The canonical commutation relation yields 
$[c T_n, P_n] = [T_n, H^{\rm clock}_n] = i\hbar$. 
Then, the clock states are orthogonal $\la \tau |\tau'\ra=\delta(\tau-\tau')$ and 
 satisfy the covariance relation $|\tau+\tau'\ra=U_C(\tau')|\tau\ra$. 
The conditional probability  (\ref{conditional-prob}) becomes
\beqa
{\rm Prob}[T_A=\tau_A|T_B=\tau_B]=\frac{\int dt~{\rm tr}[E_A(\tau_A)\rho_A(t)]{\rm tr}[E_B(\tau_B)\rho_B(t)]}{\int dt~{\rm tr}[E_B(\tau_B)\rho_B(t)]} ,
\eeqa
where $\rho_n(t)$ is the reduced state of the internal clock degrees of freedom defined as 
\cite{Smith:2019imm}
\beqa
\rho_n(t)={\rm tr}_{\H_S\backslash\H_n^{\rm clock}}\left(e^{-iH_St/\hbar}|\psi_{S_n}\ra\la\psi_{S_n}|e^{iH_St/\hbar}\right)
\eeqa
with the trace over the complement of the clock Hilbert space. 

We assume that the fiducial states of the internal clock degrees of freedom 
are  the Gaussian wave packets centered
 at $\tau=0$ with width $\sigma$:
\beqa
|\psi_n^{\rm clock}\ra=\frac{1}{\pi^{1/4}\sigma^{1/2}}\int d\tau~ e^{-\frac{\tau^2}{2\sigma^2}}|\tau\ra\,.
\eeqa

\subsection{Gravitational Time Dilation}

In order to investigate the effect of gravity on the quantum time dilation, 
 consider a Newtonian approximation of spacetime and adopt the metric, 
$g_{00}=-\alpha^2=-(1+2\Phi(\x)/c^2),\gamma_{ij}=\delta_{ij}$, where $\Phi(\x)$ is the Newtonian gravitational potential. 
Then the Hamiltonian (\ref{hamiltonian-n}) is expanded according to the number of the inverse power of $c^2$ as
\beqa
h_n&=& \alpha m_nc^2\sqrt{\frac{\gamma^{ij}P_{ni}P_{nj}}{m_n^2c^2}+\left(1+\frac{H_n^{\rm clock}}{m_nc^2}\right)^2}\nonumber\\
&\simeq & m_nc^2+H_n^{\rm clock}+H_n^{\rm ext}+H_n^{\rm int}+O(c^{-4}),
\eeqa
where the rest-mass energy term $m_nc^2$ is a constant and can be disregarded in 
$h_n$. The external Hamiltonian $H_n^{\rm ext}$ and the interaction Hamiltonian 
$H_n^{\rm int}$ are given by
\beqa
H_n^{\rm ext}&:=&\frac{\delta^{ij}P_{ni}P_{nj}}{2m_n}+m_n\Phi_n\equiv \frac{\P_n^2}{2m_n}+
m_n\Phi_n,\label{hamiltonian:ext}\\
H_n^{\rm int}&:=&-\frac{H_n^{\rm ext}H_n^{\rm clock}}{m_nc^2}
+2\frac{m_n\Phi_n(H_n^{\rm clock}+H_n^{\rm ext})}{m_nc^2}
-\frac{(H_n^{\rm ext})^2}{2m_nc^2}
-2\frac{(m_n\Phi_n)^2}{m_nc^2}\,,
\eeqa 
where $\Phi_n:=\Phi(\x_n)$. 

The reduced state of the internal clock is then given by 
\beqa
\rho_n(t)&=&{\rm tr}_{\H_S\backslash\H_n^{\rm clock}}\left[
e^{-iH_St/\hbar}|\psi_{S_n}\ra\la\psi_{S_n}|e^{iH_St/\hbar}\right]\nonumber\\
&=&\bar\rho_n(t)-it~ {\rm tr}_{\rm ext}\left(
[H_n^{\rm int}, \bar\rho_n^{\rm ext}(t)\otimes \bar\rho_n(t)]+O((H_n^{\rm int} t)^2)
\right)\nonumber\\
&=&\bar\rho_n(t)+it \left(\frac{\la H_n^{\rm ext}\ra}{m_nc^2} -2\frac{\la \Phi_n\ra}{c^2}\right)
[H_n^{\rm clock},\bar\rho_n(t)]+O(c^{-4}),
\eeqa
where $\bar \rho_n(t)=e^{-iH_n^{\rm clock}t/\hbar}\rho_n e^{iH_n^{\rm clock}t/\hbar}$ and 
$\bar \rho_n^{\rm ext}(t)=e^{-iH_n^{\rm ext}t/\hbar}\rho_n^{\rm ext} e^{iH_n^{\rm ext}t/\hbar}$.
The conditional probability  (\ref{conditional-prob}) is evaluated to leading relativistic order as
 \beqa
{\rm Prob}[T_A=\tau_A|T_B=\tau_B]=
\frac{e^{-\frac{(\tau_A-\tau_B)^2}{2\sigma^2}}}{\sqrt{2\pi}\sigma}\left[1+
\left(\frac{\la H_A^{\rm ext}\ra}{2m_Ac^2}-\frac{\la H_B^{\rm ext}\ra}{2m_Bc^2}-\frac{\la\Phi_A\ra}{c^2}+\frac{\la\Phi_B\ra}{c^2}\right)\left(1-\frac{\tau_A^2-\tau_B^2}{\sigma^2}\right)
\right] ,
\eeqa
where $\la H_n^{\rm ext}\ra=\la\psi^{\rm ext}_n|H_n^{\rm ext}|\psi_n^{\rm ext}\ra$. 
Then the average proper time read by clock $A$ conditioned on clock $B$ indicating 
the time $\tau_B$  is  
\beqa
\la T_A\ra&=&\int d\tau~{\rm Prob}[T_A=\tau|T_B=\tau_B]\tau\nonumber\\
&=&\tau_B\left(1-
\left(\frac{\la H_A^{\rm ext}\ra}{m_Ac^2}-2\frac{\la\Phi_A\ra}{c^2}\right)+\left(\frac{\la H_B^{\rm ext}\ra}{m_Bc^2}-2\frac{\la\Phi_B\ra}{c^2}\right)
\right)\,.
\label{timedelay:formal}
\eeqa
Since the external Hamiltonian is defined by Eq. (\ref{hamiltonian:ext}), $\la T_A\ra$ is written in a more suggestive form as
\beqa
\la T_A\ra=
\tau_B\left(1-
\left(\frac{\la \P_A^{2}\ra}{2m_A^2c^2}-\frac{\la\Phi_A\ra}{c^2}\right)+\left(\frac{\la \P_B^{2}\ra}{2m_B^2c^2}-\frac{\la\Phi_B\ra}{c^2}\right)
\right)\,.
\label{timedelay:final}
\eeqa
This is one of the main results of this paper and represents 
the quantum analog of time dilation formula in the Newtonian gravity.

\section{Gravitational Time Dilation of Quantum Clocks}
\label{sec3}

To evaluate the time dilation, 
for simplicity, we consider a one-dimensional problem 
in the vertical direction and  assume that the gravitational potential is approximated by 
\beqa
\Phi=gx\,,
\eeqa
where $g$ is the gravitational acceleration.  $x$ is the vertical coordinate from the surface of the Earth so that 
the external time coordinate $t$ corresponds to the proper time of an 
observer at rest on the Earth.  
Here it is to be noted that unlike the case with the external Hamiltonian 
being $\P_n^2/2m_n$ \cite{Smith:2019imm}, 
the averages of $P_n^2$ and $x_n$ evolve in time 
since $P_n$ does not commute with $H^{\rm ext}_n$,  (\ref{hamiltonian:ext}).


The external Hamiltonian (\ref{hamiltonian:ext}),  
$H_n^{\rm ext}=P_n^2/2m_n+m_ngx_n$, can be derived from the Lagrangian  $L_n$  
\beqa
L_n=\frac12 m_n\dot x_n^2-m_ngx_n .
\eeqa
Since $L_n$ does not depend on the external time $t$ explicitly,  
$H_n^{\rm ext}$ is a constant. Moreover, there  are two additional conserved quantities (Noether charges) associated with  the invariance of $L_n$ (up to total derivative). 
(i) Noether charge $Q_G$ associated with the Galilean transformation, s.t. $x_n\rightarrow x_n+vt$,
\beqa
Q_G=tP_n-m_n x_n+\frac12 m_n gt^2,
\eeqa
and (ii) Noether charge $Q_x$ associated with the spatial translation, s.t. $x_n\rightarrow x_n+a$
\beqa
Q_x=P_n+m_n gt .
\eeqa
These three conserved quantities facilitate expressing the time dilation 
at the external time $t$ for a general quantum state $|\psi_n^{\rm ext}\ra$.  
For example, since the expectation value of $H_n^{\rm ext}$ is conserved, 
the expectation value evaluated at $t$, $\la H^{\rm ext}_n\ra_t$,  is the same as its initial value: 
\beqa
\la H^{\rm ext}_n\ra_t=\frac{\la P_n^2\ra_t}{2m_n}+m_n g\la x_n\ra_t=\la H^{\rm ext}_n\ra_0=\frac{\la P_n^2\ra_0}{2m_n}+m_n g\la x_n\ra_0.
\label{average:pt}
\eeqa
Similarly, we have
\beqa
\la Q_G\ra_t&=&t\la P_n\ra_t-m_n\la x_n\ra_t+\frac12 m_n gt^2=-m_n\la x_n\ra_0 ,\\
\la Q_x\ra_t&=&\la P_n\ra_t+m_ngt=\la P_n\ra_0 .
\eeqa
{}From these, we obtain
\beqa
\la x_n\ra_t&=&\la x_n\ra_0+\frac{\la P_n\ra_0}{m_n}t-\frac12 gt^2\label{average:xt},\\
\la P_n\ra_t&=&\la P_n\ra_0-m_n gt .
\eeqa
Namely, the evolution of the average of the position is identical to 
the evolution of a classical particle with the initial position $\la x_n\ra_0$ and 
the initial velocity $\la P_n\ra_0/m_n$. 
In this sense, the trajectory is independent of the mass 
of the clock particle. This may be regarded as the weak equivalence 
principle for quantum particles.   \cite{Anastopoulos:2017jik} observed that 
the probability distribution of the position for a free-falling clock is the same as 
that of a free clock with the shift of its mean which is independent of its mass.

Plugging Eq. (\ref{average:pt}) and Eq. (\ref{average:xt}) 
into Eq. (\ref{timedelay:final}),  
the observed average time dilation between two clocks is given by
\beqa
\la T_A\ra&=&\tau_B\left(1-
\left(\frac{\la \P_A^{2}\ra}{2m_A^2c^2}-\frac{\la\Phi_A\ra}{c^2}\right)+\left(\frac{\la \P_B^{2}\ra}{2m_B^2c^2}-\frac{\la\Phi_B\ra}{c^2}\right)
\right)\nonumber\\
&=&
\tau_B\left(1-
\left(\frac{\la \P_A^{2}\ra_0}{2m_A^2c^2}-\frac{\la \P_B^{2}\ra_0}{2m_B^2c^2}\right)
+\frac{g}{c^2}\left(\la x_A\ra_0+2\frac{\la P_A\ra_0}{m_A}t-\la x_B\ra_0-2\frac{\la P_B\ra_0}{m_B}t\right)
\right) .
\label{timedelay:classical:p}
\eeqa
Note that a $\frac12 gt^2$ term disappears because it is independent of mass 
(the universality of free fall). 
It turns out that 
the differences of the initial average position and velocity between the
two clocks will contribute to the gravitational effect on the time dilation. 
This is another main result of this paper. This time dilation formula holds for a 
general quantum state, while an evaluation of the time dilation for Gaussian wave packets is given in Appendix~\ref{app:Gaussian}. 
The time dilation in general depends on the external time $t$ as well as on the initial average of the position and the momentum squared of each clock.   
Only if the average of the initial velocity for each clock is equal, 
i.e., $\la P_A\ra_0/m_A = \la P_B\ra_0/m_B$, the time dilation between
the two clocks depends only on the initial
average position and is independent of the external time $t$, that is,
the duration of the experiment.

\subsection{Time Dilation for a Clock in a Superposition}

However, for a state in a superposition, the time 
dilation would involve a term different from the classical time dilation. 
In order to investigate the effect of the superposition of states, 
we consider two clocks $A$ and $B$ and  suppose that initially 
clock $A$ is  prepared 
as a superposition of two Gaussian wave packets with average momenta $\overline{ p}_A$ 
and $\overline{ p}_A'$ with spread $\Delta_A$ and with 
mean positions $\overline{x}_A$ and $\overline{x}_A'$ 
generalizing the situation considered in \cite{Smith:2019imm,Khandelwal:2019mem} 
:
\beqa
\la P_A|\psi^{\rm ext}_A\ra=\frac{1}{N^{1/2}(\pi \Delta_A^2)^{1/4}}\left(
\cos\theta e^{-i\frac{\overline{x}_A({P_A}-\overline{ p}_A)}{\hbar}}e^{-\frac{({P_A}-\overline{ p}_A)^2}{2\Delta_A^2}}+\sin\theta e^{i\phi}e^{-i\frac{\overline{x}_A'({P_A}-\overline{ p}_A')}{\hbar}}e^{-\frac{({P_A}-\overline{p}_A')^2}{2\Delta_A^2}}\right)\,,
\label{eq:superposition}
\eeqa 
where $\theta\in[0,\pi/2)$ and $\phi\in [0,\pi]$ is 
the relative phase, and
\beqa
N=1+\sin 2\theta\exp\left(-\frac{(\overline{p}_A-\overline{ p}_A')^2}{4\Delta_A^2}-\frac{(\overline{x}_A-\overline{ x}_A')^2}{4\sigma_{xA}^2}\right)\cos\left(\phi-\frac{(\overline{x}_A+\overline{x}_A')(\overline{p}_A-\overline{p}_A')}{2\hbar}\right)
\eeqa
is a normalization factor  and $\sigma_{xA}=\hbar/\Delta_A$. 
Further, clock $B$ is  prepared in a Gaussian wave packet with average momentum 
$\overline{ p}_B$ and spread $\Delta_B$ and with mean position $\overline{x}_B$. 

Then, the average time read by $A$ conditioned on $B$ is given by 
\beqa
\la T_A\ra&=&\tau_B\Bigg{\{}1-\frac{1}{Nc^2}\left[
\left(
\frac{(\overline{ p}_A-m_Agt)^2}{2m_A^2}-g\left(\overline{x}_A+\frac{\overline{ p}_A t}{m_A}\right)
\right)\cos^2\theta+\left(
\frac{(\overline{ p}_A'-m_Agt)^2}{2m_A^2}-g\left(\overline{x}_A'+\frac{\overline{ p}_A' t}{m_A}\right)
\right)\sin^2\theta\right.\nonumber\\
&&\left.+\frac{\Delta_A^2}{4m_A^2}+\frac12 g^2t^2
\right]
-\frac{\sin 2\theta}{Nc^2} e^{-\frac{(\overline{ p}_A-\overline{ p}_A')^2}{4\Delta_A^2}
-\frac{(\overline{x}_A-\overline{ x}_A')^2}{4\sigma_{xA}^2}}
\Biggl{[}
\Biggl{(}
\frac{\left(\frac{\overline{ p}_A+\overline{ p}_A'}{2}-m_Agt\right)^2}{2m_A^2}
-\frac{\Delta_A^2(\overline{x}_A-\overline{x}_A')^2}{8m_A^2\sigma_{xA}^2}
+\frac{\Delta^2_A}{4m_A^2}\nonumber\\
&&
-\frac12 g(\overline{x}_A+\overline{x}_A')
-\frac{gt}{2m_A}(\overline{ p}_A+\overline{ p}_A')
+\frac12 g^2t^2
\Biggr{)}\cos\left(\phi-\frac{(\overline{x}_A+\overline{x}_A')(\overline{p}_A-\overline{p}_A')}{2\hbar}\right)\nonumber\\
&&
-\left(\frac{\hbar(\overline{x}_A-\overline{x}_A')}{2m_A^2\sigma_{xA}^2}
\left(\frac{\overline{ p}_A+\overline{ p}_A'}{2}-m_Agt\right)
-g\frac{\hbar t(\overline{x}_A-\overline{x}_A')}{2m_A\sigma_{xA}^2}
+g\frac{\hbar(\overline{ p}_A-\overline{ p}_A')}{2\Delta_A^2}
\right)\sin\left(\phi-\frac{(\overline{x}_A+\overline{x}_A')(\overline{p}_A-\overline{p}_A')}{2\hbar}\right)
\Biggr{]}\nonumber\\
&&+\frac{ (\overline{p}_B-m_Bgt)^2}{2m_B^2c^2}+\frac{\Delta_B^2}{4m_B^2c^2}-
\frac{g}{c^2}\left(-\frac12 gt^2+\overline{x}_B+
\frac{\overline{p}_Bt}{m_B}\right)
\Bigg{\}}\,.
\label{quantum:dilation}
\eeqa
Note that for the state given by (\ref{eq:superposition}) 
each initial expectation value, 
$\la P_A \ra_0$, $\la P_A^2 \ra_0$, and $\la x_A \ra_0$, 
is evaluated in Appendix~\ref{app:superposition}.
It coincides with the expression given in \cite{Smith:2019imm} 
when putting $g=0$ and $\overline{x}_A=\overline{x}_A'=\overline{x}_B=0$.  
It also reproduces the result by 
\cite{Khandelwal:2019mem} when putting $\phi=0$ and $\overline{p}_A=\overline{p}_A'$. 
The terms proportional to $\sin2\theta$ represent a genuin  
quantum time dilation effect. 

To make the effect of quantum time dilation manifest, we divide the time 
dilation formula  Eq. (\ref{quantum:dilation}) into the classical dilation part $K_C$ and the quantum dilation part $K_Q$ as 
$\la T_A\ra=\tau_B(1-K_{C}-K_Q)$.  $K_C$ is given by 
 the contribution of a statistical mixture of the wave packets with momentum 
 $\overline{p}_A$ and the position $\overline{x}_A$ with weight $\cos^2\theta$ and $\overline{p}_A'$ and $\overline{x}_A'$ with weight $\sin^2\theta$:  
 \beqa
 K_C&=&\frac{(\overline{ p}_A-m_A gt)^2}{2m_A^2c^2}\cos^2\theta+\frac{(\overline{ p}_A'-m_A gt)^2}{2m_A^2c^2}\sin^2\theta-\frac{(\overline{ p}_B-m_B gt)^2}{2m_B^2c^2}+
 \frac{\Delta_A^2}{4m_A^2c^2}-\frac{\Delta_B^2}{4m_B^2c^2}
 \nonumber\\
&& -\frac{g}{c^2}\left(\left(\overline{x}_A+\frac{\overline{ p}_A t}{m_A}\right)\cos^2\theta+
\left(\overline{x}_A'+\frac{\overline{ p}_A' t}{m_A}\right)\sin^2\theta-
\overline{x}_B- \frac{\overline{ p}_Bt}{m_Bc^2}\right)
\label{eq:Kc_gaussian}
 \eeqa
 and $K_Q$ is the rest of the time dilation and is given by
 \beqa
 K_Q=\frac{\sin2\theta}{8m_A^2c^2N}e^{-\frac{(\overline{ p}_A-\overline{ p}_A')^2}{4\Delta_A^2}
-\frac{(\overline{x}_A-\overline{ x}_A')^2}{4\sigma_{xA}^2}}\Biggl{\{}
\Biggl{[}2\left(\overline{p}_A'^2-\overline{p}_A^2+4m_Agt(\overline{p}_A-\overline{p}_A')+2gm_A^2(\overline{x}_A-\overline{x}_A')\right)\cos 2\theta\nonumber\\
-(\overline{p}_A-\overline{p}_A')^2-\frac{\Delta_A^2}{\sigma_{xA}^2}(\overline{x}_A-\overline{x}_A')^2
\Biggr{]}
\cos\left(\phi-\frac{(\overline{x}_A+\overline{x}_A')(\overline{p}_A-\overline{p}_A')}{2\hbar}\right)\nonumber\\
-\left[\frac{2\hbar}{\sigma_{xA}^2}(\overline{p}_A+\overline{p}_A'-4m_Agt)(\overline{x}_A-\overline{x}_A')+\frac{4\hbar m_A^2g}{\Delta_A^2}(\overline{p}_A-\overline{p}_A')
\right]\sin\left(\phi-\frac{(\overline{x}_A+\overline{x}_A')(\overline{p}_A-\overline{p}_A')}{2\hbar}\right) 
 \Biggr{\}} .
 \eeqa 
 Positive $K_Q$ implies the enhanced time dilation. 
We evaluate $K_Q$ for two cases: 
 (1) the superposition in momentum space:  $\overline{x}_A=\overline{x}_A'=0$ 
 and $\overline{p}_A\neq \overline{p}_A'$, and  
(2) the superposition in position space : $\overline{x}_A\neq \overline{x}_A'$ and $\overline{p}_A=\overline{p}_A'$. 

\subsubsection{ Momentum Superposition}

For the superposition in momentum space,  
$K_Q$ is given by
\begin{equation}
 \begin{split}
  K_Q=\frac{\sin2\theta 
}{8m_A^2c^2N}e^{-\frac{(\overline{ p}_A-\overline{p}_A')^2}{4\Delta_A^2}}
\biggl\{
\left[
2(\overline{ p}_A'^2-\overline{ p}_A^2+4 m_Agt(\overline{ p}_A-\overline{ p}_A'))\cos2\theta -
(\overline{ p}_A-\overline{ p}_A')^2
\right]\cos\phi \\
-\frac{4\hbar m_A^2g}{\Delta_A^2}(\overline{ p}_A-\overline{ p}_A')\sin\phi
\biggr\} .
 \end{split}
\end{equation}
The coefficients of $\cos\phi$ and $\sin\phi$ involve the effect of $g$. 
The coefficient of $\cos\phi$  depends linearly on $t$, while 
the coefficient of $\sin\phi$ depends only on $g$ and is independent of 
the external time $t$. 

However, the coefficient of $\sin\phi$ is negligibly small for an atomic particle. 
For  an atomic particle, $\sigma_{xA}\sim 10^{-10}{\rm m}$ and 
$m_A\sim 10^{-25}{\rm kg}$, and  $\Delta_A/m_Ac=\hbar/(m_Ac \sigma_{xA})\sim 10^{-8}$. 
Therefore, $\hbar g(\overline{p}_A-\overline{p}_A')/\Delta_A^2c^2
\sim (\hbar /\Delta_A) (g/c^2)\sim g\sigma_{xA}/c^2\sim 10^{-26}$. 
On the other hand, $((\overline{p}_A-\overline{p}_A')/m_Ac)^2\sim (\Delta_A/m_Ac)^2\sim 10^{-16}$.

\begin{figure}[htp]
	\centering
	\includegraphics[width=0.85\textwidth]{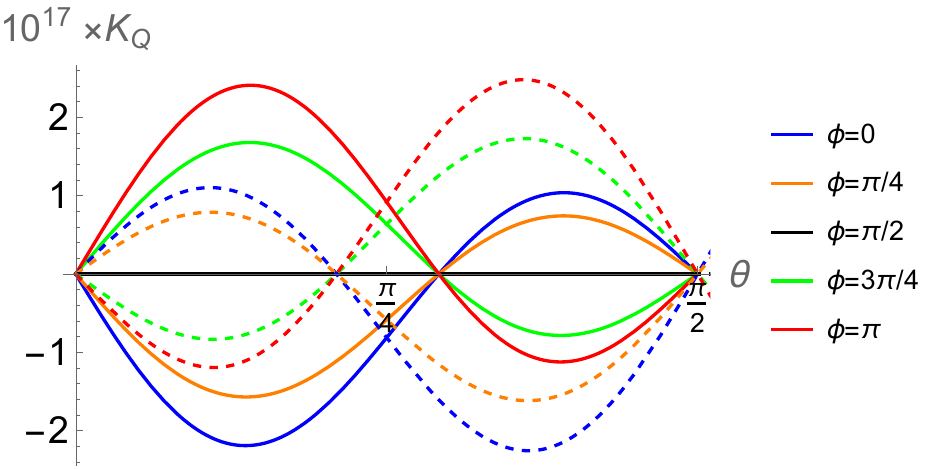}
	\caption{ \label{fig1}
	$K_Q$ for momentum superposition in a gravitational background (solid curves) and 
	in a non-gravitational background (dashed curves). 
	Curves are for $\phi=0$ (blue), 
	$\phi=\pi/4$ (orange), $\phi=\pi/2$ (black),  $\phi=3\pi/4$ (green) and $\phi=\pi$ (red). 
	 For $\phi=\pi/2$, $K_Q$ is of order $O(10^{-27})$.
	 } 
\end{figure}

To compute $K_Q$ for the momentum superposition, as in 
\cite{Smith:2019imm}, 
consider two clock particles to be $~^{87}{\rm Rb}$ atoms with a mass of $m=1.4\times 10^{-25} {\rm kg}$ and atomic radius of $\sigma_{xA}=\hbar/\Delta_A=2.5 \times 10^{-10} {\rm m}$. 
Suppose these clock particles are moving at average velocities of $\overline{v}_A'=\overline{p}_A'/m_A=15~ {\rm m/s}$
 and $\overline{v}_A=\overline{p}_A/m_A=5~ {\rm m/s}$. We further assume  $t=1~ {\rm s}$. 
Maintaing a superposition of wavepackets during 
$t\sim 1{\rm s}$ is achieved in \cite{2015Natur.528..530K}.  
 The strength of the quantum time dilation effect $K_Q$ for several $\phi$ 
 as a function of $\theta$ is shown in Fig. \ref{fig1}. 

$K_Q$ with gravity has  generically opposite sign to $K_Q$ without gravity. 
The magnitude of $K_Q$ with gravity for $\theta<\pi/4$ is larger than  $K_Q$ 
without gravity.  
The quantum time dilation effect, $\sim 10^{-17}{\rm s}$ for $t\sim 1{\rm s}$, 
  can be measured with state-of-the-art optical lattice clocks   
whose frequency measurement uncertainty recently reaches less than $10^{-20}$ \cite{Bothwell:2021fqe}, 
although the effect for $\phi=\pi/2$ is still unmeasurably small. 
For $\theta=\pi/8$,  $K_Q$ ($K_Q$ without $g$)   
 is $-2.2\times 10^{-17}$ ($1.1\times 10^{-17}$), $-1.6\times 10^{-17}$ ($7.7\times 10^{-18}$), $2.0\times 10^{-27}$ ($0$), 
 $1.7\times 10^{-17}$ ($-8.1\times 10^{-18}$), $2.4\times 10^{-17}$ ($-1.2\times 10^{-17}$) for $\phi=0,\pi/4,\pi/2,3\pi/4,\pi$, respectively.

\subsubsection{Spatial Superposition}

For the superposition in position space, $K_Q$ is given by
\begin{equation}
 \begin{split}
  K_Q = \frac{\sin2\theta 
}{8m_A^2c^2N}e^{-\frac{(\overline{ x}_A-\overline{ x}_A')^2}{4\sigma_{xA}^2}}\left\{
\left[4m_A^2g(\overline{ x}_A-\overline{ x}_A')\cos2\theta -\frac{\Delta_A^2}{\sigma_{xA}^2}(\overline{x}_A-\overline{x}_A')^2\right]\cos\phi\right.
\\
\left.\quad-\frac{4\hbar(\overline{ p}_A-2m_Agt)}{\sigma_{xA}^2}(\overline{ x}_A-\overline{ x}_A')\sin\phi
\right\} .
 \end{split}
\end{equation}
Because of the exponential suppression factor 
$\exp(-(\overline{x}_A-\overline{x}_A')^2/4\sigma_{xA}^2)$ in $K_Q$, the quantum 
interference effect is possible only when  two wave packets overlap. 
The coefficients of $\cos\phi$ and $\sin\phi$ involve the effect of $g$, and this time  
the coefficient of $\cos\phi$ is independent of $t$, while 
the coefficient of $\sin\phi$ depends linearly on $t$. 
However, the effect of $g$ in the coefficient of $\cos\phi$ is negligibly 
small for an atomic particle.   
The first term in $K_Q$ is of order $g(\overline{x}_A-\overline{x}_A')/c^2\sim g\sigma_{xA}/c^2\sim 10^{-26}$, 
while $(\Delta_A/m_Ac)^2((\overline{x}_A-\overline{x}_A')/\sigma_{xA})^2\sim (\Delta_A/m_Ac)^2\sim 10^{-16}$. The observability of the 
 quantum interference effect through interference patterns is also discussed in 
 \cite{Paczos:2022teu}. 

\begin{figure}[htp]
	\centering
	\includegraphics[width=0.8\textwidth]{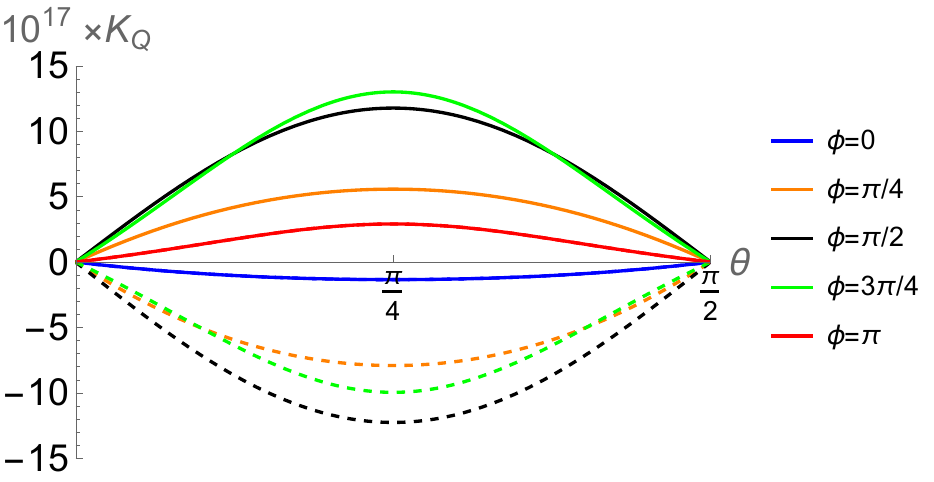}
	\caption{ \label{fig2}
	$K_Q$ for spatial superposition in a gravitational background (solid curves) and 
	in a non-gravitational background (dashed curves). 
	Curves are for $\phi=0$ (blue), 
	$\phi=\pi/4$ (orange), $\phi=\pi/2$ (black),  $\phi=3\pi/4$ (green) and $\phi=\pi$ (red).
	 For $\phi=0,\pi$, the effect of $g\neq 0$ is negligible and two curves overlap.  
	 } 
\end{figure}

$K_Q$ for the position superposition is shown in Fig. \ref{fig2}. Here we assumed  
$\overline{v}_A=\overline{p}_A/m_A=10~ {\rm m/s}, \overline{x}_A-\overline{x}_A'=2\sigma_{xA}$ and $t=1 {\rm s}$.  
For $\phi=0,\pi$, 
solid curves ($g\neq 0$) coincide with  dotted curves ($g=0$) because 
the effect of $g\neq 0$ is negligible. 
For $\theta=\pi/4$,  $K_Q$ ($K_Q$ without $g$)   
 is $-1.3\times 10^{-17}$ ($-1.3\times 10^{-17}$), $5.6\times 10^{-17}$ 
 ($-7.9\times 10^{-17}$), $1.2\times 10^{-16}$ ($-1.2 \times 10^{-16}$), 
 $1.3\times 10^{-16}$ ($-1.0\times 10^{-16}$), $2.9\times 10^{-17}$ ($2.9\times 10^{-17}$) for $\phi=0,\pi/4,\pi/2,3\pi/4,\pi$, respectively.   
It is seen that the sign of $K_Q$ with gravity is generically positive and  
is opposite to $K_Q$ without gravity for $0<\phi<\pi$, which implies that the quantum interference effect with gravity facilitates the time dilation of clocks. 



 \subsubsection{The Effect of Gravity in Quantum Interference }
 
 We have seen that the effect of nonzero $g$ in the quantum time dilation can be 
 potentially observed for clocks  
 prepared in a superposition either in momentum space or in position space.  
 In both cases, the measurable contributions originate from the initial
 expectation value of momentum, $\la P_A \ra_0$, in Eq.~(\ref{timedelay:classical:p}).
 Moreover, we can maximize the effect of $g$ on $K_Q$ by minimizing 
 the effect of $g$ on $K_C$. 
 This is achieved by arranging $\overline{p}_A$ and 
 $\overline{p}_A'$ so that the center of mass velocity of clock $A$ coincides with the velocity of clock $B$\footnote{For the case without gravity,  $K_C$ can be made to vanish 
 by arranging $\overline{p}_A$ and 
 $\overline{p}_A'$  so that the center of mass kinetic energy of clock $A$ consides with the kinetic energy of clock $B$ \cite{Smith:2019imm}. }:   
 \beqa
 \frac{\overline{p}_A\cos^2\theta+\overline{p}_A'\sin^2\theta}{m_A}
 = \frac{\overline{p}_B}{m_B} .
 \eeqa
 Then the time-dependent terms proportional to $gt$ in Eq.~(\ref{eq:Kc_gaussian})
disappear and $K_C$ becomes independent of time: 
  \beqa
 K_C=\frac{\overline{ p}_A^2}{2m_A^2c^2}\cos^2\theta+\frac{\overline{ p}_A'^2}{2m_A^2c^2}\sin^2\theta-\frac{\overline{ p}_B^2}{2m_B^2c^2}+
 \frac{\Delta_A^2}{4m_A^2c^2}-\frac{\Delta_B^2}{4m_B^2c^2}
 -\frac{g}{c^2}\left(\overline{x}_A\cos^2\theta+\overline{x}_A'\sin^2\theta-
\overline{x}_B\right), 
 \eeqa
 while $K_Q$ involves a term proportional to $gt$. Therefore, the effect of $g$ in $K_Q$ 
 would manifest as the time dependence of $K_Q$. 
 In principle, the longer the duration of the experiment, the more
  the gravitational effect is enhanced.

\section{Summary}
\label{sec4}

In this paper, extending the proper time observable proposed by \cite{Smith:2019imm} 
to a weak gravitational field, 
we derived a formula of the average proper time read by one 
clock conditioned on another clock reading a different proper 
time [Eq. (\ref{timedelay:final})].  The time dilation measured by these quantum clocks 
has the same form as that in classical relativity. 

 By considering a linear gravitational potential,    we found for a general quantum state 
that the evolution of the average of the position of a quantum clock is identical to 
the evolution of a classical particle falling under the gravitational force 
with the initial position $\la x_n\ra_0$ and 
the initial velocity $\la P_n\ra_0/m_n$, which may be regarded as 
the quantum analog of the weak equivalence principle. 
We also derived  the time dilation between clocks at the external time $t$ 
Eq. (\ref{timedelay:classical:p}), which holds  not only 
for Gaussian wave packets but also for general quantum states. 

We then considered the case in which the state of one clock is in a 
superposition of Gaussian wave packets. We found that the effect arising from quantum interference appears in the time dilation. The time dilation is of the order of 
   $10^{-17} {\rm s}$ for superpositions of wavepacket during about $1~ {\rm s}$.
The effect of gravity on 
such quantum time dilation can be potentially observed for optical lattice clocks  
 prepared in a superposition either in momentum space or in position space. 
 We also identified the condition which maximizes the effect of gravity 
 in the quantum time dilation.

\section*{Acknowledgments}
This work is supported by JSPS Grant-in-Aid for Scientific Research Number 
22K03640 (TC), No.16K17704(S.K.), and in part by Nihon University. 


\appendix

\section{Time Dilation for Wave Packets}
\label{app:Gaussian}

Here we give the time dilation for Gaussian wave packets. 

First, we consider the situation where the external degrees of freedom of two clocks 
are  initially prepared in a Gaussian state 
 localized 
around an average momentum $\overline{ p}_n$ with 
spread $\Delta_n$ and around an average position $\overline{x}_n$ with 
spread $\sigma_{xn}=\hbar/\Delta_n$ generalizing  
the situation considered in  \cite{Smith:2019imm,Khandelwal:2019mem}. 
\cite{Smith:2019imm}  considered wave packets localized in momentum space and 
\cite{Khandelwal:2019mem} considered wave packets localized in position space. 

The external state in the momentum representation 
is then given by
\beqa
\la P_n|\psi^{\rm ext}_n\ra=\frac{1}{(\pi\Delta_n^2)^{1/4}}
e^{-i\frac{\overline{x}_n({P_n}-\overline{ p}_n)}{\hbar}}
e^{-\frac{({P_n}-\overline{ p}_n)^2}{2\Delta_n^2}}\,,
\label{wavepacket:p}
\eeqa
and the initial average of $P_n^2$ is then given by
\beqa
\la P_n^2\ra_{0}=\la \psi^{\rm ext}_n|P_n^2|\psi^{\rm ext}_n\ra
=\int dp ~p^2|\la p|\psi^{\rm ext}_n\ra |^2=\overline{ p}_n^2+\frac{\Delta_n^2}{2}\,.
\label{average:p}
\eeqa
The external state in the position representation is given by
\beqa
\la x_n|\psi^{\rm ext}_n\ra&=&\int dp \la x_n|p\ra\la p|\psi^{\rm ext}\ra\nonumber\\
&=&\frac{1}{(\pi \sigma_{xn}^2)^{1/4}}e^{\frac{i}{\hbar}x_n\overline{ p}_n}e^{-\frac{(x_n-\overline{x}_n)^2}{2\sigma_{xn}^2}}\,,
\label{wavepacket:x}
\eeqa 
where $\sigma_{xn}=\hbar/\Delta_n$, and  
the initial average of the position is  $ \overline{x}_n$, 
$\la x_n \ra_{0}=\int dx ~x|\la x|\psi^{\rm ext}_n\ra |^2=\overline{x}_n$. 

Plugging $\la \P_n^{2}\ra_0$ 
into Eq. (\ref{timedelay:classical:p}),  
the time dilation between two clocks is given by
\beqa
\la T_A\ra
&=&\tau_B\left(1-
\left(\frac{\la \P_A^{2}\ra_0}{2m_A^2c^2}-\frac{\la \P_B^{2}\ra_0}{2m_B^2c^2}\right)
+\frac{g}{c^2}\left(\la x_A\ra_0+2\frac{\la P_A\ra_0}{m_A}t-\la x_B\ra_0-2\frac{\la P_B\ra_0}{m_B}t\right)
\right)\nonumber\\
&=&
\tau_B\left[1-\left(\frac{(\overline{ p}_A-m_Agt)^2}{2m_A^2c^2}
-\frac{(\overline{ p}_B-m_Bgt)^2}{2m_B^2c^2}
+\frac{\Delta_A^2}{4m_A^2c^2}
-\frac{\Delta_B^2}{4m_B^2c^2}\right)
+\frac{g}{c^2}\left(\overline{x}_A+\frac{\overline{ p}_At}{m_A}
-\overline{x}_B-\frac{\overline{ p}_Bt}{m_B}\right)
\right]\,.
\label{timedelay:wavepacket}
\eeqa
Therefore, supposing $\Delta_A/m_A=\Delta_B/m_B$, the time dilation between 
two quantum clocks agrees with the  classical relativistic time dilation.

\section{Initial expectation values for a superposition of states}
\label{app:superposition}

We summarize the expressions of expectation values of momentum, momentum squared, and
position operators for a superposition of Gaussian wave
packets given by Eq.~(\ref{eq:superposition}):
\begin{align}
    \langle P_A \rangle 
    &=
    \overline{ p}_A \cos^2\theta 
    + \overline{p}_A' \sin^2\theta
    - \frac{\sin 2\theta}{N}\, 
    e^{
    - \frac{(\overline{p}_A^2 - \overline{p}_A')^2}{4\Delta_A^2}
    - \frac{(\overline{x}_A - \overline{x}_A')^2}{4\sigma_{xA}^2}
    } 
 \nonumber
    \\
    & \qquad
    \times \left[
    \frac{(\overline{p}_A - \overline{p}_A')\cos 2\theta}{2}
    \cos 
    \left(\phi - 
    \frac{(\overline{x}_A + \overline{x}_A')(\overline{p}_A - \overline{p}_A')}{2\hbar}
    \right)   
    + \hbar\frac{\overline{x}_A - \overline{x}_A'}{2\sigma_{xA}^2}
    \sin 
    \left(\phi - 
    \frac{(\overline{x}_A + \overline{x}_A')(\overline{p}_A - \overline{p}_A')}{2\hbar}
    \right)
    \right]
    ,
\\
    \langle P_A^2 \rangle 
    &= \overline{p}_A^2 \cos^2\theta + \overline{p}_A'^2 \sin^2\theta 
    + \frac{\Delta_A^2}{2}
    - \frac{\sin 2\theta}{N} \,
    e^{
    - \frac{(\overline{p}_A^2 - \overline{p}_A')^2}{4\Delta_A^2}
    - \frac{(\overline{x}_A - \overline{x}_A')^2}{4\sigma_{xA}^2} }
 \nonumber
    \\
    & \quad
    \times
    \Biggl\{
    \left[
    \frac{\overline{p}_A^2 - \overline{p}_A'^2}{2} \cos 2\theta
    + \left(\frac{\overline{p}_A - \overline{p}_A'}{2}\right)^2
    + \Delta_A^2 \left(\frac{\overline{x}_A - \overline{x}_A'}{2\sigma_{xA}}\right)^2
    \right]
    \cos \left(\phi - 
    \frac{(\overline{x}_A + \overline{x}_A')(\overline{p}_A - \overline{p}_A')}{2\hbar}
    \right)
    \nonumber \\
    & \qquad
    + \hbar\frac{(\overline{p}_A + \overline{p}_A')(\overline{x}_A - \overline{x}_A')}{2\sigma_{xA}^2}
    \sin \left(\phi - 
    \frac{(\overline{x}_A + \overline{x}_A')(\overline{p}_A - \overline{p}_A')}{2\hbar}
    \right)
    \Biggr\}
    , 
\\
    \langle x_A \rangle 
    &= \overline{x}_A \cos^2\theta + \overline{x}_A' \sin^2\theta
    - \frac{\sin 2\theta}{N}
    e^{
    - \frac{(\overline{p}_A^2 - \overline{p}_A')^2}{4\Delta_A^2}
    - \frac{(\overline{x}_A - \overline{x}_A')^2}{4\sigma_{xA}^2} }
 \nonumber
 \\
    &\quad 
    \times 
 \left[
    \frac{\overline{x}_A - \overline{x}_A'}{2} 
    \cos 2\theta
    \cos \left(\phi - 
    \frac{(\overline{x}_A + \overline{x}_A')(\overline{p}_A - \overline{p}_A')}{2\hbar}
    \right)
    - \hbar\frac{\overline{p}_A - \overline{p}_A'}{2\Delta_A^2}
    \sin \left(\phi - 
    \frac{(\overline{x}_A + \overline{x}_A')(\overline{p}_A - \overline{p}_A')}{2\hbar}
    \right)
 \right]
    .
\end{align}
Note that the terms proportional to $\frac{\sin 2\theta}{N}$ in the
above expectation values represent contributions of quantum interference.

\bibliographystyle{apsrev4-1}
\bibliography{references}

\end{document}